\documentclass[aps,prd,onecolumn,superscriptaddress,showpacs,floatfix,10pt,nofootinbib]{revtex4}
\usepackage{amssymb}
\usepackage{textcomp}
\usepackage{graphicx}
\usepackage{epsfig}
\usepackage{amsmath}
\usepackage{slashed}
\usepackage{units}
\usepackage{hyperref}
\usepackage{array}
\usepackage{lmodern}
\usepackage{booktabs}
\usepackage{float}

\begin{document}

\title{Sub-GeV-scale signatures of hidden braneworlds up to the Planck scale\\ in a $SO(3,1)$-broken bulk}
\author{Coraline Stasser}
\affiliation{Laboratory of Analysis by Nuclear Reactions, Department of Physics,
University of Namur, 61 rue de Bruxelles, B-5000 Namur, Belgium}
\author{Micha\"{e}l Sarrazin}
\email{michael.sarrazin@ac-besancon.fr}
\thanks{Corresponding author}
\affiliation{Institut UTINAM, CNRS/INSU, UMR 6213, Universit\'{e} Bourgogne--Franche-Comt%
\'{e}, 16 route de Gray, F-25030 Besan\c con Cedex, France}
\affiliation{Lyc\'{e}e Saint-Paul, 8 Boulevard Diderot, F-25000
Besan\c con, France}
\affiliation{Laboratory of Analysis by Nuclear Reactions, Department of Physics,
University of Namur, 61 rue de Bruxelles, B-5000 Namur, Belgium}

\begin{abstract}
Many-brane universes are at the heart of several cosmological scenarios related to physics beyond the Standard Model. It is then a major concern to constrain these approaches. Two-brane Universes involving $SO(3,1)$-broken 5D bulks are among the cosmological models of interest. They also allow considering matter exchange between branes, a possible way to test these scenarios. Neutron disappearance (reappearance) toward (from) the hidden brane is currently tested with high-precision experiments to constrain the coupling constant $g$ between the visible and hidden neutron sectors. When dealing with the sub-GeV-scale quantum dynamics of fermions, any pair of braneworlds can be described by a non-commutative two-sheeted space-time $M_4\times Z_2$ from which $g$ emerges. 
Nevertheless, the calculation of the formal link between $g$ for a neutron and $SO(3,1)$-broken 5D bulks remains an open problem until now although necessary to constrain these braneworld scenarios. Thanks to a phenomenological model, we derive $g$ -- for a neutron -- between the two braneworlds endowed with their own copy of the standard model in a $SO(3,1)$-broken 5D bulk. Constraints on interbrane distance and brane energy scale (or brane thickness) are discussed. While brane energy scale below the GUT scale is excluded, energy scale up to the Planck limit allows neutron swapping detection in forthcoming experiments.
\end{abstract}

\maketitle

\section{Introduction}

The desert -- i.e. no new physics in colliders between the TeV and the GUT
scales \cite{desert1,desert2,desert3,desert4} -- is a realistic but disappointing scenario feared in
the context of the recent LHC results \cite{LHC1,LHC2}. However, some
high-precision experiments at low energy could detect signatures of GUT
scale and beyond. For instance, neutron electric dipole moment (nEDM) \cite%
{nEDM1,nEDM2,nEDM3} or proton decay \cite{protondecay1,protondecay2,protondecay3} are such signatures. As a
consequence, prospecting for new low-energy tests of physics beyond the
Standard Model is of crucial interest. It offers an alternative and
additional route to colliders to probe new physics with experiments at quite
low cost. Moreover, many works \cite%
{domainwall,first1,first2,origin,E8E8,branevsDW1,branevsDW2,branevsDW3,branevsDW4,RSI,RSII,ExtraDBranreview,
ekpy,pyro,branedark,6D1,6D2,6D3,6D4,6D5,6D6,review1,review2,review3,review4,M4xZ2,pheno,npm,npmth,exp,ftl,DGS1,DGS2,twogen,ChungFreese,BWED}
describe our visible Universe -- our visible world -- as a domain wall \cite%
{domainwall,branevsDW1,branevsDW2,branevsDW3,branevsDW4} (i.e. a 3-brane) inside a higher dimensional bulk,
generally with five \cite{RSI,RSII,ExtraDBranreview,
ekpy,pyro,DGS1,DGS2,twogen,ChungFreese,BWED} or sometimes six dimensions \cite{6D1,6D2,6D3,6D4,6D5,6D6}.
Some models also consider that many braneworlds could coexist within the
bulk \cite%
{E8E8,RSI,ekpy,pyro,branedark,review1,review2,review3,review4,M4xZ2,pheno,npm,npmth,exp,ftl,ChungFreese}%
. The interest for such many-world scenarios traces back to the Ho\v{r}%
ava-Witten \cite{E8E8} approach linking the $E_{8}\times E_{8}$ heterotic
super-string theory in ten dimensions to eleven-dimensional supergravity on
the orbifold $R_{10}\times S_{1}/Z_{2}$. Assuming a six-dimensional
compactification on a Calabi-Yau manifold, this model leads to a $%
M_{4}\times S_{1}/Z_{2}$ Universe (i.e. a five-dimensional bulk with a
compactified extra dimension) in which ordinary- and hidden-sector particles
live on two different 3-branes (or braneworlds) located at each boundary of
the orbifold. Along this line of thought, many models emerge such as the
Randall and Sundrum's (RS) solution of the hierarchy problem \cite{RSI}, or
the so-called ekpyrotic \cite{ekpy} universes as alternatives to cosmic
inflation or to explain dark matter and dark energy \cite{branedark, review1,review2,review3,review4}%
. These frameworks have also been extended to a non-compact fifth dimension
in $M_{4}\times R_{1}$ or $M_{4}\times R_{1}/Z_{2}$ bulks \cite{RSI,RSII,ExtraDBranreview,ekpy,pyro,DGS1,DGS2,twogen,ChungFreese} where the bulk
metric can be warped or not \cite{RSI,DGS1,DGS2}. In many scenarios, while our
visible braneworld sustains particles of the Standard Model (the TeV brane
in RS-like models), the hidden braneworld (or Planck brane) supports a
Planck-scale physics. In some other approaches, the hidden world should
possess its own copy of the Standard Model \cite%
{review1,review2,review3,review4,ChungFreese,CF2,CF3,manyfold,ApCF,manyCF}. This is the case in the
two-brane Universe scenarios where the bulk curvature violates the $SO(3,1)$
isometry thus leading to a violation of Lorentz invariance on branes \cite%
{review1,review2,review3,review4,ChungFreese,CF2,CF3}. The typical metric is then the Chung-Freese
metric such that \cite{ChungFreese,CF2,CF3}: 
\begin{eqnarray}
ds^{2} &=&g_{AB}^{(5)}dx^{A}dx^{B}\text{, with }A,B=0,1,...,4  \label{metric}
\\
&=&g_{\mu \nu }^{(4)}dx^{\mu }dx^{\nu }-dz^{2}\text{, with }\mu ,\nu
=0,1,...,3,  \notag
\end{eqnarray}%
where $g_{\mu \nu }^{(4)}=diag(1,-a^{2}(z),-a^{2}(z),-a^{2}(z)).$ In the
following, when we refer to $x$, it denotes the usual 4D space-time
components while $z$ is the fifth extra dimension along which the branes are
located. At a cosmological scale this approach offers a non-inflationary
solution to the cosmological horizon problem \cite%
{review1,review2,review3,review4,ChungFreese,CF2,CF3,manyfold,ApCF,manyCF}. In addition, Lorentz
non-invariant terms lead to an attractive phenomenology which can be tested 
\cite{nEDM3,lorentz}. As a consequence, it is a major concern to
further explore experimental expectations and constraints on these
braneworlds scenarios with $SO(3,1)$-broken isometry, either at a
cosmological scale or in particle physics. Since these scenarios contain a
copy of the Standard Model in each brane, they should allow for matter
exchange between both branes, even at low-energy physics. This is the
phenomenology studied in the present paper.

High-energy phenomena are often considered when dealing with braneworld
phenomenology. However, in the last decade, one of us (M.S.) and
collaborators showed theoretically that matter exchange could occur at low
energy between closest braneworlds in the bulk \cite%
{M4xZ2,pheno,npm,npmth,exp,ftl}. For instance, neutron swapping is allowed
between our visible Universe and a parallel one hidden in the bulk \cite%
{M4xZ2,pheno,npm,npmth,exp}. Such an effect is probed in some experiments
involving search for passing-through-walls neutrons in the vicinity of a
nuclear reactor \cite{npm,npmth,murmur} or search for unusual leaks when
dealing with ultra-cold neutron storage \cite{exp}. The ability for a
neutron to leap from our visible world into a hidden braneworld (or
reciprocally) is given by a swapping probability $p$ \cite{pheno}, which can
be experimentally probed \cite{npm,npmth,exp,murmur}.

The study of this phenomenology is made possible by the fact that any
Universe with two braneworlds -- regardless of the underlying model -- is
equivalent to an effective non-commutative two-sheeted space-time $%
M_{4}\times Z_{2}$ when following the dynamics of particles at low energies
below the GeV scale \cite{M4xZ2}. Although many braneworlds can coexist
within the bulk, in a first approximation, one can consider a two-brane
Universe consisting of two copies of the Standard Model, localized in two
adjacent 3-branes. While these two branes are mutually invisible to each
other at the zeroth-order approximation, matter fields in separate branes
mix at the first-order approximation mainly through $\mathcal{L}_{c}=ig%
\overline{\psi }_{+}\gamma ^{5}\psi _{-}+ig\overline{\psi }_{-}\gamma
^{5}\psi _{+}$, where $\psi _{\pm }$ are the Dirac fermionic fields in each
braneworld -- denoted $(+)$ and $(-)$ \cite{M4xZ2,pheno}. $g$ is the
two-brane or interbrane coupling constant and can be related to the swapping
probability such that $p\propto g^{2}$ \cite{pheno,npm,npmth,exp}.

Fundamentally, branes in the bulk can be described by $\xi $-thick domain
walls (with a brane energy scale $M_{B}\sim 1/\xi $ ), which are kink or
anti-kink solitons of scalar fields allowing fermions -- or gauge bosons --
trapping on the 3D world \cite{domainwall,branevsDW1,branevsDW2,branevsDW3,branevsDW4}. Surprisingly, the
equivalence between two-brane models and the non-commutative two-sheeted
space-time approach is general and neither relies on the domain walls
features or on the bulk properties (dimensionality, number of compact
dimensions, and metric) \cite{M4xZ2}. As a consequence, in the $M_{4}\times
Z_{2}$ model, the fundamental parameter $g$ is a black box which implicitly
contains the bulk properties, the distance $d$ between branes in the bulk,
but also their thickness $\xi $ and some properties of the states localized
on each brane \cite{M4xZ2}.

In the first derivation of the $M_{4}\times Z_{2}$ model of a two-brane
Universe \cite{M4xZ2}, it was shown that $g$ can be explicitly computed
against brane and bulk parameters \cite{M4xZ2} when dealing with a
domain-wall description of branes. But for an illustrative purpose, and for
the sake of simplicity, a simple scenario was considered \cite{M4xZ2}, with
a flat $M_{4}\times R_{1}$ bulk with fermions trapped along two thick domain
walls. In this naive scenario the mass of the fermions on each brane was
only related to Kaluza-Klein states -- the interbrane coupling could then be
expressed by \cite{M4xZ2}: 
\begin{equation}
g\sim (1/\xi )\exp (-d/\xi ).  \label{g0}
\end{equation}%
Here, it is worth noting that the brane energy scale could be as high as the
Planck scale, i.e. well beyond the reach of direct searches at high energy
with particle colliders. However, depending on the ratio $d/\xi $, $g$ can
still reach values permitting to explore the braneworld through matter
swapping at low energy \cite{pheno,npm}. This motivating result is also at
the origin of the present paper.

To go further, considering the current experimental context \cite%
{pheno,npm,npmth,exp} involving neutron experiments, it is crucial to
explicitly derive $g$ for a neutron, which is not a Kaluza-Klein state. In
addition, regarding the models here under consideration, where the bulk
possesses a $SO(3,1)$-broken isometry, one must also be able to consider
warped metrics. Nevertheless, a full \textit{ab initio} computation of $g$
is out of reach such as and for now. Indeed, strictly speaking, a full
description of the neutron in the domain wall approach would need for a
valid description of the Standard Model on the domain-wall branes and to
numerically solve the equations giving birth to the neutron. For instance, a
lattice Quantum Chromodynamics (QCD) on a Universe with two domain-wall
branes would be necessary, a huge work far beyond the scope of the present
paper. As a consequence, before anything else, a phenomenological approach
is first necessary to identify the main features of $g$ for a neutron
against the bulk metric and dimensionality, the brane energy scale, and
interbrane distance. This is the purpose of the present paper. In section %
\ref{framework}, we recall the main results and technical difficulties of
the framework which motivated the present work. In section \ref{fp}, we
introduce the new phenomenological approach which allows us to address the
problem of the derivation of $g$ for the neutron for two thick branes in a $%
SO(3,1)$-broken bulk. At last, in section \ref{coupling}, constraints on
interbrane distance and brane energy scale (or brane thickness) are
considered according to existing experimental bounds on $g$, which are also
discussed in the context of future experiments.

\section{$M_{4}\times Z_{2}$ low-energy limit of a universe with two
domain-wall 3-branes: main results and open problems}

\label{framework}

As introduced here above, whatever the high-energy theory of a two-brane
Universe (i.e. whatever the number or properties of bulk scalar fields
responsible for particle trapping on branes, the number of extra dimensions
or the bulk metric, etc.), the fermion dynamics on both branes at low energy
corresponds to the dynamics of fermions in a $M_{4}\times Z_{2}$ space-time
in the context of the non-commutative geometry \cite{M4xZ2}. Let us recall some fundamentals of this framework. For an
illustrative purpose, one first considers two braneworlds described by two
topological defects in the bulk. For instance, we can consider two domain
walls corresponding to a kink -- anti-kink pair of solitons in a $%
M_{4}\times R_{1}$ flat 5D bulk. A simple Lagrangian for such a system is:

\begin{equation}
\mathcal{L}_{M_{4}\times R_{1}}=\frac{1}{2}\left( \partial _{A}\Phi \right)
\left( \partial ^{B}\Phi \right) -V(\Phi )+\overline{\Psi }\left( i\Gamma
^{A}\left( \partial _{A}+i\mathcal{A}_{A}\right) -\lambda \Phi \right) \Psi ,
\label{L1}
\end{equation}%
where $\mathcal{A}_{A}$ is a $U(1)$ bulk gauge field and $\Phi $ is the
scalar field. The potential $V(\Phi )$ is assumed to ensure the existence of
kink-like solutions, i.e. of domain walls, by following the
Rubakov-Shaposhnikov concept \cite{domainwall}. $\Psi $ is the massless
fermionic matter field. $\Psi $ is coupled to the scalar field $\Phi $
through a Yukawa coupling term $\lambda \overline{\Psi }\Phi \Psi $ with $%
\lambda $ the coupling constant.

In our previous work \cite{M4xZ2}, it was shown that $\mathcal{L}%
_{M_{4}\times R_{1}} $ reduces to the effective Lagrangian $\mathcal{L}%
_{M_{4}\times Z_{2}} $ for energies below the GeV scale. Then, the effective
phenomenological discrete two-point space $Z_{2}$ replaces the continuous
real extra dimension $R_{1}$. At each point along the discrete extra
dimension $Z_{2}$, there is a four-dimensional space-time $M_{4}$ endowed
with its own metric. Each $M_{4}$ sheet describes each braneworld considered
as being separated by a phenomenological distance $\delta =1/g$. This result
is obtained from an approach inspired by the construction of molecular
orbitals in quantum chemistry, here extended to fermionic bound states on
branes. Then, $g$ is proportional to an overlap integral of the fermionic
wave functions of each $3$-brane over the extra dimension $R_{1}$ \cite%
{M4xZ2}.

The effective $M_{4}\times Z_{2}$ Lagrangian for the fermion dynamics in a
two-brane Universe is \cite{M4xZ2,pheno}: 
\begin{equation}
\mathcal{L}_{M_{4}\times Z_{2}}\sim \overline{\Psi }\left( {i{\slashed{D}}%
_{A}-M}\right) \Psi .  \label{L2}
\end{equation}%
Labelling $(+)$ (respectively $(-)$) our brane (respectively the hidden
brane), we write: $\Psi =\left( 
\begin{array}{c}
\psi _{+} \\ 
\psi _{-}%
\end{array}%
\right) $ where $\psi _{\pm }$ are the wave functions in the branes $(\pm )$
and: 
\begin{equation}
{i{\slashed{D}}_{A}-M}=\left( 
\begin{array}{cc}
i\gamma ^{\mu }(\partial _{\mu }+iqA_{\mu }^{+})-m & ig\gamma ^{5}-im_{r} \\ 
ig\gamma ^{5}+im_{r} & i\gamma ^{\mu }(\partial _{\mu }+iqA_{\mu }^{-})-m%
\end{array}%
\right) .  \label{Dirac}
\end{equation}%
$A_{\mu }^{\pm }$ are the electromagnetic four-potentials on each brane $%
(\pm )$. $m$ is the mass of the bound fermion on a brane. The mass mixing
term, due to the off-diagonal mass term $m_{r}$, results from the two-domain
wall Universe \cite{M4xZ2}. The phenomenology related to $m_{r}$ can be
neglected when compared to the phenomenology related to $g$ as shown
elsewhere \cite{pheno} and as briefly recalled in section \ref{subscoup}.
The derivative operators acting on $M_{4}$ and $Z_{2}$ are $D_{\mu }=\mathbf{%
1}_{8\times 8}\partial _{\mu }$ ($\mu =0,1,2,3$) and$\ D_{5}=ig\sigma
_{2}\otimes \mathbf{1}_{4\times 4}$, respectively, and the Dirac operator
acting on $M_{4}\times Z_{2}$ is defined as ${\slashed{D}=}\Gamma
^{N}D_{N}=\Gamma ^{\mu }D_{\mu }+\Gamma ^{5}D_{5}$ where: $\Gamma ^{\mu }=%
\mathbf{1}_{2\times 2}\otimes \gamma ^{\mu }$\ and\ $\Gamma ^{5}=\sigma
_{3}\otimes \gamma ^{5}$. $\gamma ^{\mu }$ and $\gamma ^{5}=i\gamma
^{0}\gamma ^{1}\gamma ^{2}\gamma ^{3}$ are the usual Dirac matrices and $%
\sigma _{k}$ ($k=1,2,3$) the Pauli matrices. Note that Eq. (\ref{Dirac}) is
typical of non-commutative $M_{4}\times Z_{2}$ two-sheeted space-times \cite%
{M4xZ2}. We refer to the terms proportional to $g$ as geometrical mixing.

Regarding the electromagnetic field, it was shown \cite{M4xZ2} that the
five-dimensional $U(1)$ bulk gauge field must be substituted by an effective 
$U(1)_{+}\otimes U(1)_{-}$ gauge field in the $M_{4}\times Z_{2}$
space-time. The Dvali-Gabadadze-Shifman mechanism \cite{DGS1,DGS2} leads to the
gauge field localization on the branes and $U(1)_{\pm }$ are the gauge
groups of the photon fields on each brane: the bulk gauge field $\mathcal{A}%
_{A}$ splits into $A_{\mu }^{\pm }$. The electromagnetic field $\slashed{A}%
\sim diag(iq\gamma ^{\mu }A_{\mu }^{+},iq\gamma ^{\mu }A_{\mu }^{-})$ is
then introduced in the Dirac equation through ${\slashed{D}}_{A}\rightarrow {%
\slashed{D}}+\slashed{A}$ \cite{M4xZ2}.

In this approach \cite{M4xZ2}, it must be noted that the mass $m$
corresponds to a Kaluza-Klein state \cite{M4xZ2} and $g$ is given by Eq. (%
\ref{g0}). So, in order to address the behaviour of the neutron, we should
first consider massless fermionic Kaluza-Klein states. Next, many fermionic
fields should be introduced in the model. Not only to get quark families,
but also because two fermionic fields are required at least to get both
chirality states on the brane when considering the fundamental Kaluza-Klein
states \cite{twogen}. Of course, gauge and Higgs fields should be also
included in the domain-wall-brane model to dress massless states to retrieve
the expected standard masses on branes. Indeed, the Higgs field contributes
to quark masses while, considering the neutron, its mass mainly arises from
the strong interaction between quarks. Of course, one should also ensure to
retrieve the Standard Model symmetry as well as the efficiency of the
mechanisms needed to confine each field on each brane \cite{branevsDW1,branevsDW2,branevsDW3,branevsDW4} when
dealing with the bulk metric. There is possibly many ways to reach these
goals which meet the landscape problem encountered in superstring models 
\cite{string}. Anyway, this issue is then far beyond the scope of the
present paper. As a consequence, a more relevant and phenomenological
approach must be considered for now to address the main features of the
coupling constant $g$ when dealing with neutron in a two-brane Universe with
a given bulk metric. This is the mainspring of the approach introduced in
the next section.

\section{$M_{4}\times Z_{2}$ low-energy limit of a universe with two
D3-branes: an approach for the neutron phenomenology}

\label{fp}

\begin{figure}[th]
\centering
\includegraphics[width=6.0 cm]{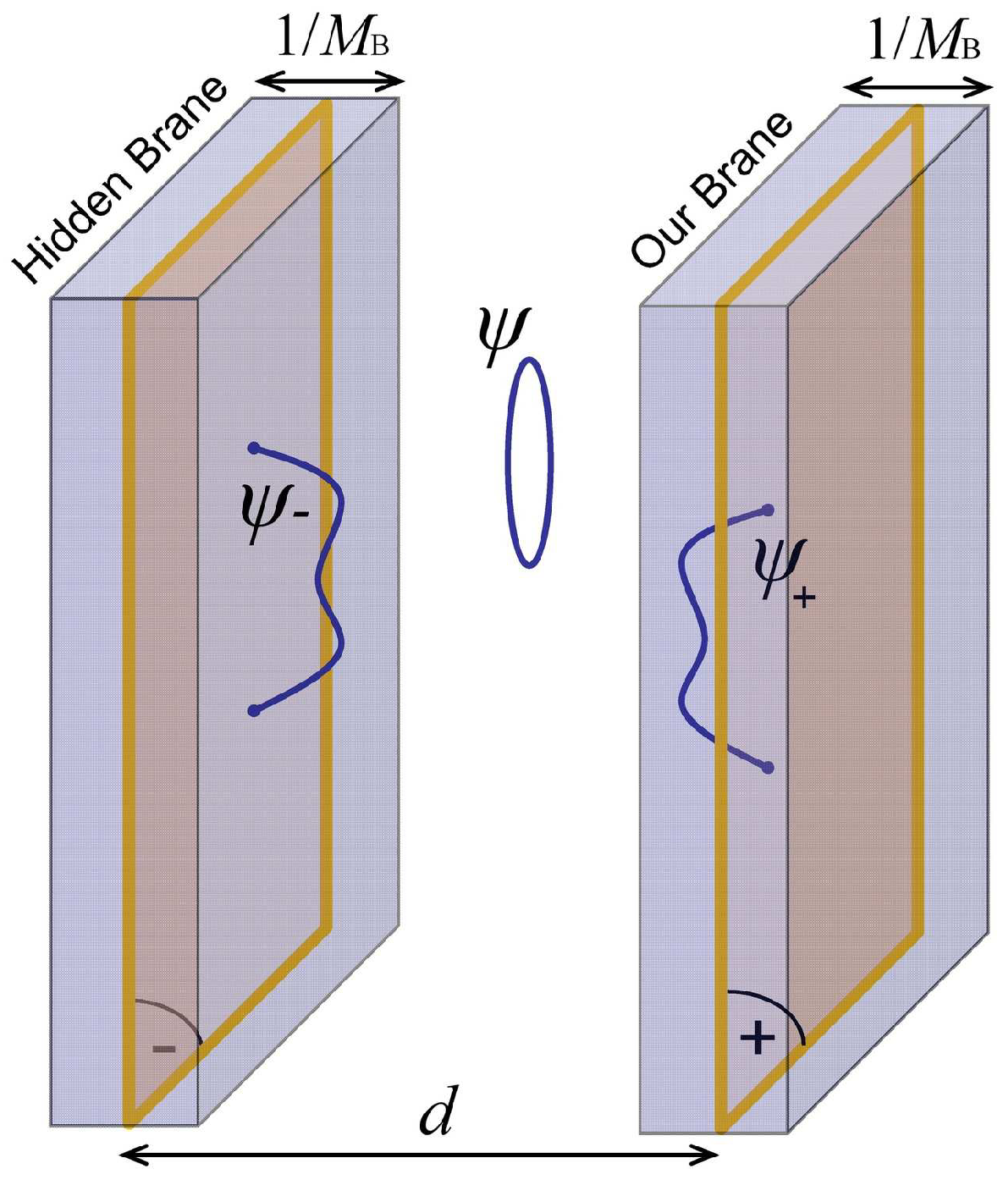}
\caption{(Color online). Sketch of the problem under study. A
string-inspired approach allows treating the problem from a phenomenological
point of view by introducing brane quark fields $\protect\psi _{\pm }$ and a
bulk quark-like field $\protect\psi $. While D3-branes (yellow) correspond
to strings' endpoints, they can be considered as thick 3-branes (bluish)
when dressed with strings. Brane thickness is introduced through the brane
energy scale $M_{B}$ which is roughly the string energy scale. Interbrane
distance $d$ in the $M_{4}\times R_{1}$ bulk is greater than $M_{B}^{-1}$.}
\end{figure}

In the present section, we derive the relevant phenomenological $M_{4}\times
Z_{2}$ low-energy limit for a universe with two D3-branes embedded in a
Chung-Freese bulk with $SO(3,1)$-broken isometry. By doing so, we seek to
estimate the interbrane coupling constant $g$ for a neutron -- i.e. a 
\textit{udd} quark triplet -- in such a universe. While the fermion - hidden
fermion coupling is the core of our attention, bosons' coupling between both
branes is a correction which can be here neglected \cite{M4xZ2,pheno}
without loss of generality. In the following, we will study the coupling
between quarks' fields of each brane. We consider a simplified\footnote{%
For instance, in the present approach, Kalb-Ramond and Ramond-Ramond fields
are not under consideration \cite{string}.} string-inspired \cite{string}
phenomenological model (Fig. 1). We thus introduce two quark field families,
each one located in each brane (like the open strings on a brane) and bulk
fermionic fields free to propagate in the bulk (like the closed strings).
The whole problem is described by the full action $%
S=S_{bulk}+S_{brane(+)}+S_{brane(-)}+S_{coupling}$ which contains the four
terms described hereafter.

\subsection{Effective quark action on branes}

We first consider that each brane is endowed with its own copy of the
Standard Model. A quark field on a brane is then assumed as resulting from
an open string initially in its zero-mass state and fixed on the brane.
Fundamentally, a quark gets its mass from interaction with the Higgs field
on branes only. In addition, the neutron will get its mass mainly from the
strong force interaction between quarks in the \textit{udd} quark triplet.
As a consequence, one should introduce the Higgs and $SU(3)$ gauge fields on
the brane in order to compute the whole neutron properties. Nevertheless,
for instance, usual \textit{ab initio} computations of masses or magnetic
moments of hadrons use tedious numerical computation to describe
gluons/quarks dynamics from lattice QCD. Such an approach is far beyond the
scope of the present work. As a consequence, to identify the main behaviour
of the neutron in our model, we use for now the Constituent Quark Model
(CQM) \cite{CQMbook,CQM1,CQM2,CQM3}. In hadron, each idealized current quark (or naked
quark) is dressed by the gluons/virtual quarks sea surrounding it. Dressed
quarks are called constituent quarks. The effective masses of constituent
quarks can be derived from experimental data, and of course, differ from
current quarks' masses. Of course, \textit{ab initio} computations in
chromodynamics would be necessary to explain the values of the masses of the
constituent quarks. But from a phenomenological point of view, the CQM works
pretty well to calculate masses or magnetic moments \footnote{%
Effective magnetic moments $\mu _{q}$ of constituent quarks can be easily
derived from their effective masses $m$ and from quarks' charges $q$ since $%
\mu _{q}=q\hbar /2m$. Effective magnetic moments can be simply combined to
retrieve the hadron magnetic moment. The hadron mass can also simply be
retrieved by summing effective masses of each constituent quark plus a
spin-spin corrective term \cite{CQMbook,CQM1,CQM2,CQM3}.} \cite{CQMbook,CQM1,CQM2,CQM3}. In
section \ref{subscoup}, the interbrane coupling constant of the neutron will
be naturally derived from CQM \cite{CQMbook,CQM1,CQM2,CQM3}. As a consequence, in the
following, each quark on brane will be endowed with its effective
phenomenological mass $m$, which implicitly contains the effects of Higgs
field and quark-to-quark interactions on branes. Then, the quark actions
(whatever the flavour) on each brane located on $z=\pm d/2$ resume to: 
\begin{equation}
S_{brane(\pm )}=\int d^{4}x\sqrt{\left\vert g_{\pm }^{(4)}\right\vert }%
\overline{\psi }_{\pm }\left( i\Gamma _{\pm }^{\mu }\left( \partial _{\mu
}+C_{\mu }^{\pm }+iqA_{\mu }^{\pm }\right) -m\right) \psi _{\pm },
\label{Lbrane}
\end{equation}%
where $\psi _{\pm }\left( x\right) $ (respectively $g_{\pm }^{(4)}$) are the
usual 4D spinors of quark fields (respectively the induced metric tensors)
on each brane. $C_{\mu }^{\pm }$ and $\Gamma _{\pm }^{\mu }$ \ are the spin
connections and the Dirac matrices taken on the branes and deduced from the
general 5D equations (\ref{Gamma}) to (\ref{connexion}) introduced in the
next section. $m$ is the quark constituent mass induced on each brane. $%
A_{\mu }^{\pm }$ are the electromagnetic potentials on each brane. The
expression of $S_{brane(\pm )}$ would correspond to infinitely thin branes,
i.e. D3-branes. Considering string-dressed 3-branes, that means one expects
that gauge fields and fermions of the Standard Model are contained in the
thickness $\xi $ of the 3-brane, i.e. they do not spread beyond $z>\xi
=M_{B}^{-1}$. As a consequence, the brane thickness is assumed to be small
compared to the interbrane distance $d$.

\subsection{Effective quark-like fermion action in the bulk}

We assume that for each quark on branes (as an open string) there is a dual
bulk fermion able to carry flavour and charges in the bulk (as a closed
string). Since particles' interactions of the Standard Model are supposed to
exist on branes only, we assume that bulk particles are sterile in relation
to each other. The dual quark-like bulk state is assumed to be massless%
\footnote{%
Strings can have excitation states leading to effective masses, which here
are model-dependent multiple of $M_{B}$ \cite{string}. Then, we neglect the
small corrections of these states as we are mainly concerned by the order of
magnitude of $g$ in the current experimental context.} and related to a
closed string in its zero-mass state. Since we neglect quarks' interaction
in the bulk, the bulk triplet is also massless. The bulk quark-like field $%
\psi (x,z)$ follows then the action: 
\begin{equation}
S_{bulk}=\int d^{4}xdz\sqrt{\left\vert g^{(5)}\right\vert }\overline{\psi }%
\left( i\Gamma ^{A}\left( \partial _{A}+C_{A}+iq\mathcal{A}_{A}\right)
\right) \psi  \label{Lbulk}
\end{equation}%
where the Dirac matrices $\Gamma ^{A}$ follow the relationship $\left\{
\Gamma ^{A},\Gamma ^{B}\right\} =2g^{(5)AB}\mathbf{1}$. By contrast, we
define the "flat" Dirac matrices $\gamma ^{A}$ through: $\left\{ \gamma
^{A},\gamma ^{B}\right\} =2\eta ^{(5)AB}\mathbf{1}$ with $\eta ^{(5)AB}$ the
Minkowski metric with a $(+,-,-,-,-)$ signature. $\mathcal{A}_{A}$ is the
effective five dimensional electromagnetic vector potential such that $%
\mathcal{A}_{4}=0$, and which is supposed to be zero everywhere except on
braneworlds. $q\overline{\psi }\Gamma ^{A}\mathcal{A}_{A}\psi $ ensures
charge conservation through the bulk. The five dimensional Dirac matrices in
curved space-time take the form: 
\begin{equation}
\Gamma ^{A}(x,z)=e_{B}^{A}(x,z)\gamma ^{B},  \label{Gamma}
\end{equation}%
where $e_{a}^{A}$ defines the vielbein according to: 
\begin{equation}
g^{(5)AB}=e_{C}^{A}(x,z)e_{D}^{B}(x,z)\eta ^{(5)CD}.  \label{vielbein}
\end{equation}%
The spin connection must satisfy the expression: 
\begin{equation}
C_{A}(x,z)=\frac{1}{4}\Gamma _{B}\left[ {\partial _{A}\Gamma ^{B}+\Gamma
_{CA}^{B}\Gamma ^{C}}\right] ,  \label{connexion}
\end{equation}%
where $\Gamma _{CA}^{B}$ are the Christoffel symbols for the metric field
under consideration.

\subsection{Effective bulk -- brane fields coupling action}

The fermion fields on each brane can be considered as source/well terms for
the fermionic bulk field. The coupling between the brane fields and the bulk
fields occurs naturally as a mass coupling on branes, and we get:%
\begin{eqnarray}
S_{coupling} &=&-\int d^{4}xdz\sqrt{\left\vert g^{(4)}\right\vert }
\label{Lcoupling} \\
&&\times \left\{ \frac{m}{M_{B}^{1/2}}\left( \overline{\psi }_{+}\psi +%
\overline{\psi }\psi _{+}\right) \delta (z-d/2)\right.  \notag \\
&&\left. +\frac{m}{M_{B}^{1/2}}\left( \overline{\psi }_{-}\psi +\overline{%
\psi }\psi _{-}\right) \delta (z+d/2)\right\} .  \notag
\end{eqnarray}%
From the point of view of the domain-wall approach, $\psi $ and $\psi _{\pm
} $ are fundamentally the same fields \cite{M4xZ2}, this is why we use the
simple mass mixing implying $m$, as the coupling occurs only on each brane.
Now, though the expression of $S_{brane(\pm )}$ corresponds to infinitely
thin branes, $S_{coupling}$ introduces the finite thickness $\xi \sim
1/M_{B} $ of the branes along which the coupling occurs (due to the spatial
extent of strings). In $S_{coupling}$, the power $1/2$ of $M_{B}$ ensures
the correct dimensionality of the problem.

\subsection{Validity domain of the model}

$\bullet$ If both 3-branes are too close to each other, typically $d\leq
1/M_{B}$, direct interactions between fermion fields of each 3-brane could
occur and exotic fields related to open strings stretched between both
3-branes could appear \cite{string}. In such a case, both 3-branes should be
then considered as a single 3D world, with visible and dark sectors. The
last is a component which can be added to the Standard Model to restore some
symmetries \cite{Mirror1,Mirror2,Mirror3}. This has been studied previously in the literature
for instance in the context of mirror particle paradigm \cite{Mirror1,Mirror2,Mirror3}. This
topic is out of the scope of the present work. To ensure two independent
3-branes and so the validity of our model, we roughly assume that $d>2/M_{B}$%
.

$\bullet$ In the bulk, the neutron becomes a triplet of free but entangled
quark-like bulk states. In the bulk, due to the relativistic free motion of
bulk quarks, the triplet can stretch along usual spatial dimensions when
propagating along the extra dimension. This could prevent neutron
reappearance in a brane. Indeed, due to the finite range of the strong force
and in order to restore the neutron, quarks reappearing on a brane must not
be too distant from each other. Then, the interbrane distance must not be
too large to avoid bulk quarks to separate from each other beyond the strong
force range in a brane. As a consequence, one can claim that neutron
exchange between braneworlds is not possible if the interbrane distance $d$
exceeds roughly $0.5$ fm, i.e. the coupling constant $g$ then collapses to
zero.

\subsection{$M_{4}\times Z_{2}$ limit of the two-brane universe by
propagating equations of motion over the extra dimension}

Let us first consider the equation of motion for the bulk field from $S$.
The bulk field simply follows the relationship:%
\begin{eqnarray}
&&\sqrt{\left\vert g^{(5)}\right\vert }\left( i\Gamma ^{A}\left( \partial
_{A}+C_{A}+iq\mathcal{A}_{A}\right) \right) \psi   \label{motion1} \\
&=&\sqrt{\left\vert g^{(4)}\right\vert }\frac{m}{M_{B}^{1/2}}\psi _{+}\delta
(z-d/2)+\sqrt{\left\vert g^{(4)}\right\vert }\frac{m}{M_{B}^{1/2}}\psi
_{-}\delta (z+d/2),  \notag
\end{eqnarray}%
which is the expected $5$-dimensional Dirac equation supplemented by the
source/well terms induced by the boundary conditions on the branes. Equation
(\ref{motion1}) can be then easily propagated over the extra dimension. Due
to mass shell constraint on the branes, one imposes the condition:%
\begin{equation}
\left( i\Gamma ^{\mu }\left( \partial _{\mu }+C_{\mu }+iqA_{\mu }\right)
-m\right) \psi =0  \label{shell}
\end{equation}%
and Eq. (\ref{motion1}) becomes:%
\begin{eqnarray}
&&\left( \gamma ^{5}\partial _{z}+m\right) \psi   \label{motionfree} \\
&=&\sqrt{\frac{\left\vert g_{+}^{(4)}\right\vert }{\left\vert
g_{+}^{(5)}\right\vert }}\frac{m}{M_{B}^{1/2}}\psi _{+}\delta (z-d/2)+\sqrt{%
\frac{\left\vert g_{-}^{(4)}\right\vert }{\left\vert g_{-}^{(5)}\right\vert }%
}\frac{m}{M_{B}^{1/2}}\psi _{-}\delta (z+d/2),  \notag
\end{eqnarray}%
since $C_{4}=0$ and $\Gamma ^{4}=-i\gamma ^{5}$ due to the metric choice in
Eq. (\ref{metric}). Now, we can introduce the Green function $G$ -- of the
free field $\psi $ -- which obeys to:%
\begin{equation}
\left( \gamma ^{5}\partial _{z}+m\right) G(z-z^{\prime })=\delta
(z-z^{\prime }),  \label{green}
\end{equation}%
leading to:%
\begin{eqnarray}
G(z) &=&\frac{1}{2\pi }\int \frac{i\gamma ^{5}\kappa +m}{\kappa ^{2}+m^{2}}%
e^{-i\kappa z}d\kappa   \notag \\
&=&(1/2)e^{-m\left\vert z\right\vert }\left( \mathbf{1}+sign(z)\gamma
^{5}\right) ,  \label{greenfunc}
\end{eqnarray}%
with $\gamma ^{0}G^{\dagger }(z)\gamma ^{0}=G(-z)$. Solving Eq. (\ref%
{motionfree}), $\psi $ can be simply expressed as:%
\begin{equation}
\psi (x,z)=\sqrt{\frac{\left\vert g_{+}^{(4)}\right\vert }{\left\vert
g_{+}^{(5)}\right\vert }}\frac{m}{M_{B}^{1/2}}G(z-d/2)\psi _{+}(x)+\sqrt{%
\frac{\left\vert g_{-}^{(4)}\right\vert }{\left\vert g_{-}^{(5)}\right\vert }%
}\frac{m}{M_{B}^{1/2}}G(z+d/2)\psi _{-}(x).  \label{Psibulk}
\end{equation}%
From Eqs. (\ref{Psibulk}) and (\ref{greenfunc}), one deduces that bulk field
originating from a brane, does not propagate along the extra dimension, but
has an evanescent component with a decay constant equal to $m$ along the
extra dimension\footnote{%
If massive bulk modes $M$ $\propto M_{B}$ were considered, one needs to
replace $m$ by $\left\vert m-kM_{B}\right\vert $ (where $k$ is a number
greater than $1$) in the argument of the exponential term of the bulk field
propagator. Then, for an interbrane distance greater than $1/M_{B}$, the
contributions from massive bulk modes quickly drop with a decay constant
about $kM_{B}$ and can be neglected as expected since $kM_{B}\gg m$.}. As a
consequence, despite the coupling between brane and bulk fields, quarks
remain localized on their respective brane and cannot propagate through the
bulk. Then, as shown hereafter, the only way for quarks to escape from a
brane, is to jump towards another brane thanks to a quantum-tunnelling-like
effect.

Injecting the expression of $\psi $ given by Eq. (\ref{Psibulk}) in $%
S_{coupling}$ (see Eq. (\ref{Lcoupling})), we retrieve the mass mixing and
the geometrical mixing terms found in the non-commutative $M_{4}\times Z_{2}$
approach (see section \ref{framework}):%
\begin{equation}
S_{coupling}=\int d^{4}x\left( im_{r}\overline{\psi }_{+}\psi _{-}-im_{r}%
\overline{\psi }_{-}\psi _{+}+\overline{\psi }_{+}ig\gamma ^{5}\psi _{-}+%
\overline{\psi }_{-}ig\gamma ^{5}\psi _{+}\right) ,  \label{Coupling}
\end{equation}%
where: 
\begin{equation}
g=(1/2)\frac{m^{2}}{M_{B}}\left( R^{-3/2}+R^{3/2}\right) e^{-md},\text{ }
\label{g}
\end{equation}%
and $m_{r}=g$, with $R=a_{-}/a_{+}$ ($a_{\pm }=a(z=\pm d/2)$). Note that Eq.
(\ref{g}) is invariant through $R\rightarrow R^{-1}$. In Eq. (\ref{Coupling}%
) we successively applied a convenient SU(2) rotation: 
\begin{equation}
\left( 
\begin{array}{c}
\psi _{+} \\ 
\psi _{-}%
\end{array}%
\right) \rightarrow \left( 
\begin{array}{cc}
e^{-i\pi /4} & 0 \\ 
0 & e^{i\pi /4}%
\end{array}%
\right) \left( 
\begin{array}{c}
\psi _{+} \\ 
\psi _{-}%
\end{array}%
\right) ,  \label{SU(2)}
\end{equation}%
and a convenient rescaling: 
\begin{equation}
\left( 
\begin{array}{c}
\psi _{+} \\ 
\psi _{-}%
\end{array}%
\right) \rightarrow \left( 
\begin{array}{cc}
a_{+}^{3/2} & 0 \\ 
0 & a_{-}^{3/2}%
\end{array}%
\right) \left( 
\begin{array}{c}
\psi _{+} \\ 
\psi _{-}%
\end{array}%
\right) ,  \label{resc}
\end{equation}%
to ensure a more usual wave-function normalization. In Eq. (\ref{Coupling}),
self-coupling terms of the form $\overline{\psi }_{\pm }\Gamma \psi _{\pm }$
are neglected. Indeed, $\overline{\psi }_{\pm }\Gamma \psi _{\pm
}=m^{2}/M_{B}$ thus introducing a mass correction such that $m\rightarrow
m(1+m/M_{B})$. For $m=340$ MeV \cite{CQMbook,CQM1,CQM2,CQM3} and $M_{B}$ ranging
between GUT and Planck scales, such a correction is far beyond any current
experimental accuracy \cite{LHC1,LHC2}.

Now, following the same procedure, Eq. (\ref{Lbrane}) describing quarks on
each brane becomes (after rotation (\ref{SU(2)}) and rescaling (\ref{resc})):

\begin{eqnarray}
S_{brane(\pm )} &=&\int d^{4}x\overline{\psi }_{\pm }\left( i\gamma
^{0}\left( \partial _{0}+iqA_{0}^{\pm }\right) \right.  \notag \\
&&+ia_{\pm }^{-1}\gamma ^{\eta }\left( \partial _{\eta }+iqA_{\eta }^{\pm
}\right)  \label{Lbrane2} \\
&&\left. +(3/2)\left( \partial _{z}a\right) _{\pm }a_{\pm }^{-1}\gamma
^{5}-m\right) \psi _{\pm }.  \notag
\end{eqnarray}%
In the following, the terms $(3/2)\left( \partial _{z}a\right) _{\pm }a_{\pm
}^{-1}\gamma ^{5}$ will be neglected without any loss of generality since
the Lorentz symmetry breaking introduced by these terms \cite{lorentz} can
be neglected in our present problem thanks to current experimental data \cite%
{nEDM3}.

Finally, the relevant action for the dynamics of the quark fields $\psi
_{\pm }$ on each brane is the effective $M_{4}\times Z_{2}$ action $%
S_{M_{4}\times Z_{2}}$ which is the restriction of $%
S=S_{bulk}+S_{brane(+)}+S_{brane(-)}+S_{coupling}$. Indeed, from Eq. (\ref%
{Coupling}) and Eq. (\ref{Lbrane2}), one gets $S_{M_{4}\times
Z_{2}}=S_{brane(+)}+S_{brane(-)}+S_{coupling}=\int \mathcal{L}_{M_{4}\times
Z_{2}}d^{4}x$, with:

\begin{equation}
\mathcal{L}_{M_{4}\times Z_{2}}\sim \overline{\Psi }\left( {i{\slashed{D}}%
_{A}-M}\right) \Psi ,  \label{final}
\end{equation}%
where $\Psi =\left( 
\begin{array}{c}
\psi _{+} \\ 
\psi _{-}%
\end{array}%
\right) $, and: 
\begin{equation}
{i{\slashed{D}}_{A}-M}=\left( 
\begin{array}{cc}
i\widehat{\gamma }_{\pm }^{\mu }(\partial _{\mu }+iqA_{\mu }^{+})-m & 
ig\gamma ^{5}-im_{r} \\ 
ig\gamma ^{5}+im_{r} & i\widehat{\gamma }_{\pm }^{\mu }(\partial _{\mu
}+iqA_{\mu }^{-})-m%
\end{array}%
\right) ,  \label{DiracFinal}
\end{equation}%
with $\left( \widehat{\gamma }_{\pm }^{0},\widehat{\gamma }_{\pm
}^{1,2,3}\right) =\left( \gamma ^{0},a_{\pm }^{-1}\gamma ^{1,2,3}\right) $.
When $a_{\pm }^{-1}\longrightarrow 1$, Eq. (\ref{DiracFinal}) fully matches
Eq. (\ref{Dirac}) above. It is noticeable that the present approach still
leads to a non-commutative $M_{4}\times Z_{2}$ space-time description of the
two-brane Universe \cite{M4xZ2,graphene} (see section \ref{framework}). In
fact, Eq. (\ref{final}) is simply the generalization of the $M_{4}\times
Z_{2}$ action for the Chung-Freese metric as shown in a previous work \cite%
{ftl}. Strictly speaking, the choice of such a metric does not fundamentally
change the physics of the two-brane Universe at low energy as described in
our previous papers \cite{pheno,npm,npmth,exp}. All the more, the value of $%
g $ changes against the ratio $R=a_{-}/a_{+}$ of the warp factors $a_{\pm }$
as shown by Eq. (\ref{g}). The low-energy phenomenology induced by $a_{\pm
}^{-1}\neq 1$ is fully detailed in a previous work \cite{ftl}. We just note
that the factors $a_{\pm }^{-1}$, occurring in the \textquotedblleft
usual\textquotedblright\ Dirac operator in each brane, will affect the
values of the momentum and of the kinetic energy which now differ in each
brane \cite{ftl}. Anyway, the new contributions occurring from the non-null
differences between the kinetic energies (and the momenta) of each brane are
negligible for neutrons with a kinetic energy lower or equal to that of
thermal neutrons (i.e. about $25$ meV or less) \cite{ftl}. At last,
regarding the values of $g$ (see Eq. (\ref{g})), it is not possible to
experimentally discriminate the contribution of $M_{B}$ from the warp
contribution $R$. One can then conveniently substitute the brane thickness $%
M_{B}^{-1}$ by an effective one including the warped metric effect such that 
$M_{B}^{-1}\rightarrow M_{B}^{-1}(1/2)\left( R^{-3/2}+R^{3/2}\right) $.
Anyway, $M_{B}$ remains lower or equal than $M_{Planck}$ whatever $R$.

\subsection{Phenomenology and neutron - hidden neutron coupling constant}

\label{subscoup}

Assuming the non-relativistic character of the CQM \cite{CQMbook,CQM1,CQM2,CQM3}, it is
instructive to consider the non-relativistic limit of the two-brane Dirac
equation (see Eq. (\ref{DiracFinal})). One gets a two-brane Pauli equation: $%
i\hbar \partial _{t}\Psi =\left\{ \mathbf{H}_{0}+\mathbf{H}_{cm}+\mathbf{H}%
_{c}\right\} \Psi $, with $\mathbf{H}_{0}=diag(\mathbf{H}_{+},\mathbf{H}%
_{-}) $ where $\mathbf{H}_{\pm }$ are the usual four-dimensional Pauli
Hamiltonian expressed in each braneworld, and $\Psi =\left( 
\begin{array}{c}
\psi _{+} \\ 
\psi _{-}%
\end{array}%
\right) $ where $\psi _{\pm }$ are now the Pauli spinors. Moreover, coupling
terms appear (in natural units) \cite{M4xZ2, pheno}: 
\begin{equation}
\mathbf{H}_{c}=\left( 
\begin{array}{cc}
0 & im_{r}c^{2} \\ 
-im_{r}c^{2} & 0%
\end{array}%
\right) ,  \label{Hm}
\end{equation}%
which is obviously the mass mixing term, and \cite{M4xZ2, pheno}

\begin{equation}
\mathbf{H}_{cm}=\left( 
\begin{array}{cc}
0 & -ig\widehat{\mathbf{\mu }}\cdot \left( \mathbf{A}_{+}-\mathbf{A}%
_{-}\right) \\ 
ig\widehat{\mathbf{\mu }}\cdot \left( \mathbf{A}_{+}-\mathbf{A}_{-}\right) & 
0%
\end{array}%
\right) ,  \label{Hcm}
\end{equation}%
where $\mathbf{A}_{\pm }$ are the local magnetic vector potentials in each
brane and $\widehat{\mathbf{\mu }}=\mu \mathbf{\sigma }$ is the magnetic
moment operator of the fermion. There is no pure geometrical mixing, instead 
$\mathbf{H}_{cm}$ relates to a mixed geometrical/electromagnetic coupling
involving the magnetic moment. This is allowed by the pseudo-scalar-coupling
character of the geometrical mixing. Obviously $\mathbf{H}_{cm}$ also exists
at the relativistic energy scale, like the magnetic moment, although it does
not explicitly appear in the Dirac equation. Here, the coupling strength
between fermion spinors of the visible and hidden worlds becomes clearly
dependent on the magnetic potentials which enhance the magnitude of the
geometrical mixing. Since here $m_{r}c^{2}=g\hbar c$, from Eqs. (\ref{Hm})
and (\ref{Hcm}), one can easily check that $\mathbf{H}_{cm}$ dominates $%
\mathbf{H}_{c}$ when $\left\vert \mathbf{A}_{+}-\mathbf{A}_{-}\right\vert
>A_{c}$. The critical field $A_{c}$ is given by $A_{c}$ $=2mc/q$, where $m$
is here the mass of the constituent quark, and $q$ its charge. $A_{c}\approx
7 $ Tm for the constituent quark down. This value must be compared with
these of expected astrophysical magnetic potentials about $10^{9}$ Tm \cite%
{vecpot1,vecpot2}. Then, here the geometrical/electromagnetic coupling $\mathbf{H}%
_{cm}$ would be larger than the mass mixing $\mathbf{H}_{c}$ by $8$ orders
of magnitude. As a consequence, we usually neglect $\mathbf{H}_{c}$ \cite%
{pheno}.

From the previous equations, one can show that a neutron should oscillate
between two states, one localized in our brane, the other localized in the
hidden world \cite{M4xZ2} following a similar two-brane Pauli equation%
\footnote{%
Reader will find details of the related phenomenology in our previous works 
\cite{pheno}.}. Then, assuming that $g$ (respectively $\mu$) refers here to
the coupling constant (respectively the magnetic moment) of the neutron,
using the quark constituent model \cite{CQMbook,CQM1,CQM2,CQM3}, one must verify: 
\begin{equation}
g\widehat{\mathbf{\mu }}=\sum\limits_{q}g_{q}\widehat{\mathbf{\mu }}_{q}
\label{rela}
\end{equation}%
where $g_{q}$ (respectively $\widehat{\mathbf{\mu }}_{q}$) refers to the
coupling constant (respectively the magnetic moment operator) of each quark
constituting the neutron with $\widehat{\mathbf{\mu }}=\sum\limits_{q}%
\widehat{\mathbf{\mu }}_{q}$. Since $m_{up}\approx m_{down}\approx m=340$
MeV \cite{CQMbook,CQM1,CQM2,CQM3}, one simply gets $g\approx g_{up}\approx g_{down}$.
This approach could be generalized to any chargeless baryon\footnote{%
For charged particles, the swapping between braneworlds must be dramatically
frozen \cite{npmth,exp}.} endowed with a magnetic moment. Obviously, the
large neutron lifetime and the huge number of neutrons produced in a nuclear
reactor \cite{npm,npmth,murmur} make neutron highly competitive to probe
two-brane physics by contrast to more exotic hadrons \cite{mesdec}.

\section{Bounds on interbrane distance and brane energy scale against
neutron-hidden neutron coupling}
\label{coupling}

\begin{figure}[th]
\centering
\includegraphics[width=8.9 cm]{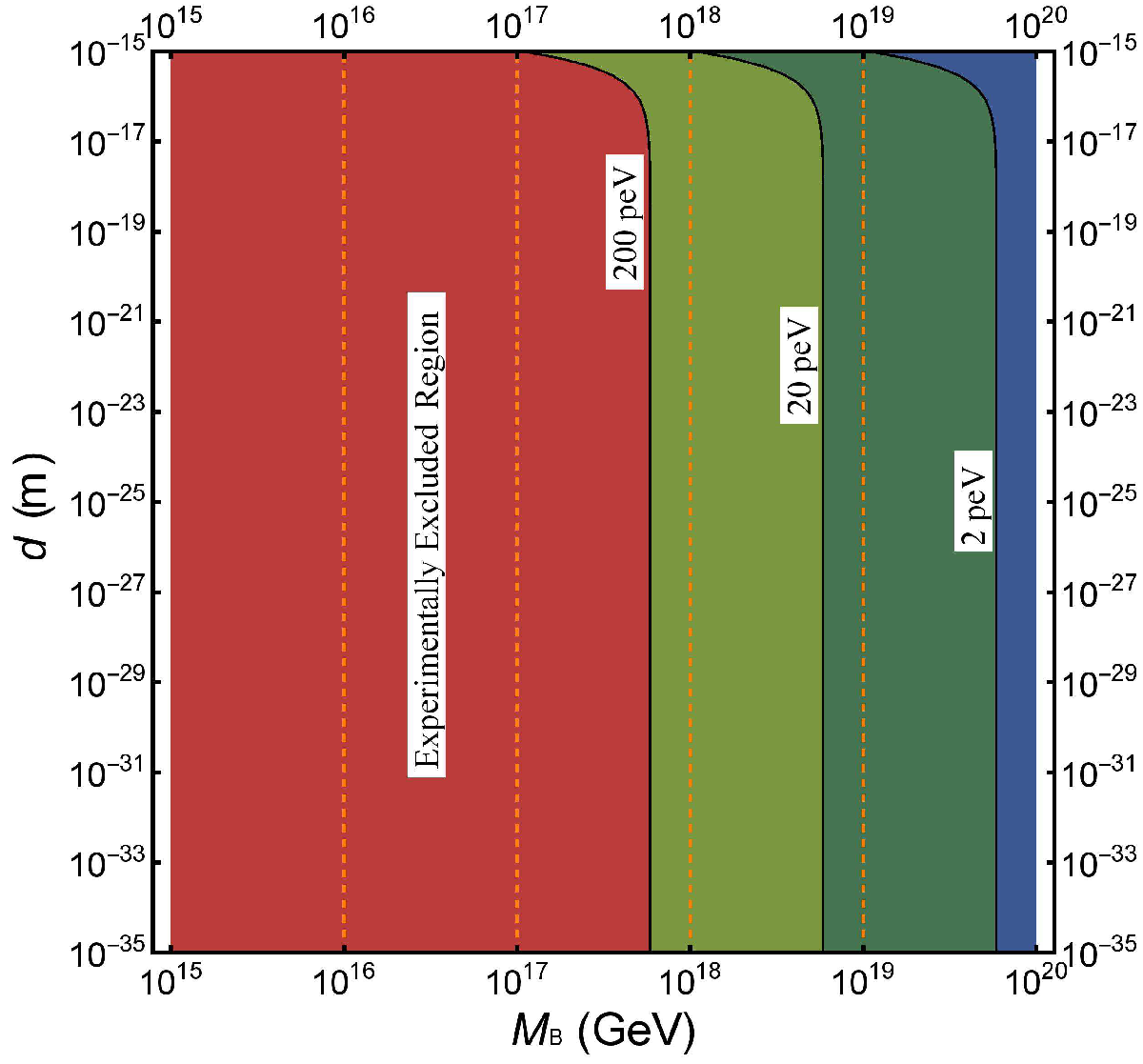}
\caption{(Color online). Bounds on the interbrane distance $d$ in a 5D bulk
and on the brane energy scale $M_B$ against the coupling constant $g$. The $%
M_B$ scale includes GUT and Planck energies ($10^{16}$ GeV and $10^{19}$ GeV
respectively). The red region for values greater than $g=200$ peV (or $%
10^{-3}$ m$^{-1}$ in natural units) is excluded with confidence from
experimental data \protect\cite{npm}. The light green region with $20 < g <
200$ peV (or $10^{-4}<g<10^{-3}$ m$^{-1}$) is expected to be likely
reachable in future experiments. The dark green region with $2 < g < 20$ peV
(or $10^{-5}<g<10^{-4}$ m$^{-1}$) will be partially reachable in future
experiments. The blue region is out of range for now. For interbrane
distances greater than $0.5$ fm, neutron exchange is supposed to be
precluded ($g=0$ m$^{-1}$) by the model.}
\end{figure}

Figure 2 shows bounds on the interbrane distance $d$ in a 5D bulk, and on
the brane energy scale $M_B$, against the coupling constant $g$. The $M_B$
scale includes GUT and Planck energies. Values greater than $200$ peV (i.e. $%
10^{-3}$ m$^{-1}$) are excluded with confidence from experimental data \cite%
{npm}. For interbrane distances greater than $0.5$ fm, neutron exchange is
supposed to be precluded ($g=0$) by the model (see section \ref{fp}), in
agreement with the exponential decay of $g$ (see Eq. (\ref{g})) as $m^{-1}
\approx 0.58$ fm. As a significant result, Fig. 2 shows that neutron
disappearance/reappearance observation would imply a brane energy scale
greater than the GUT scale. From Eq. (\ref{g}), assuming a Chung-Freese
metric, the ratio $R$ of the warp factors should not exceed $20$ to stay in
the correct energy scale domain. Regarding now the green domains in Fig. 2,
it is noteworthy that the coupling constant $g$ gets values ranging between $%
2$ and $200$ peV -- i.e. $10^{-3}<g<10^{-5}$ m$^{-1}$ -- compatible with the
observation of neutron disappearance/reappearance in the next coming
experiments \cite{npm,npmth}. If one doubts of a fine tuning of the
interbrane distance around $1/m$, and if one assumes that neutron swapping
can occur (i.e. $d$ is lower than $0.5$ fm), then short interbrane distances
should prevail leading to: $g\sim $ $m^{2}/M_{B}\approx 2.4 \times 10^{-4}$
m $^{-1}$ $\approx 50$ peV at the (reduced) Planck scale. Such a value of $g$
is very promising for next coming experiments \cite{murmur}.

\section{Conclusion}

In the context of the $M_{4}\times Z_{2}$ low-energy description of a
two-brane Universe, we have derived the explicit expression of the coupling
constant $g$ between the visible state of the neutron in our visible
braneworld and the hidden state of the neutron in the hidden braneworld.
This phenomenological approach allows studying $g$ against the interbrane
distance and the brane thickness (or brane energy scale), here in the
framework of the Chung-Freese two-brane Universes involving $SO(3,1)$-broken
5D bulks. According to current experimental bounds on $g$, we have shown
that a successful detection of the neutron swapping would lead to reject a
braneworld energy scale below the GUT energy scale. While colliders tend to
reject new physics at TeV and beyond, it is shown that even if the brane
energy scale corresponds to the Planck scale, $g$ can reach values
motivating new passing-through-walls-neutron experiments purposed to detect
neutron swapping between braneworlds \cite{npm,npmth,murmur}. While we have
focused here on a 5D bulk with a Chung-Freese metric, our approach could be
extended to a 6D bulk and beyond, compact or non-compact, warped or not.
This will be considered in further work to test the robustness of the
present results.

\section*{Acknowledgements}

This work is supported by the MURMUR collaboration
(www.murmur-experiment.eu). C.S. is supported by a FRIA doctoral grant from
the Belgian F.R.S-FNRS. The authors are grateful to Guillaume Pignol and
Christopher Smith for their useful comments on this work. The authors thank
Nicolas Reckinger for reading the manuscript.


\end{document}